\documentclass[twocolumn]{aastex61}
\usepackage{hyperref}
\usepackage{amsmath,amstext}
\usepackage[T1]{fontenc}
\usepackage[figure,figure*]{hypcap}
\usepackage{graphicx}
\usepackage{multirow,bigdelim}

\def\sun{$_{\odot}$}

\def\f28{${f}_{2-8{\rm keV}}$}

\def\sun{$_{\odot}$}

\def\ergscm2{erg s$^{-1}$ cm$^{-2}$}
\def\yr-1{yr$^{-1}$}

\def\asec{\ifmmode^{\prime\prime}\else$^{\prime\prime}$\fi}

\def\secspt{$\buildrel{\prime\prime}\over .$}

\def\Chandra{{\it Chandra}}

\received{May 10, 2018}
\submitjournal{ApJ Letters}

\shorttitle{X-ray Afterglow Fading in GW170817/GRB170817A} 
\shortauthors{Nynka {\it et al.}}
\begin{document}

\title{Fading of the X-ray Afterglow of \\ Neutron Star Merger GW170817/GRB170817A at 260 days}

\correspondingauthor{Melania Nynka}
\email{melania.nynka@mcgill.ca}

\author[0000-0002-3310-1946]{Melania Nynka}
\affil{McGill Space Institute and Department of Physics, McGill University, 3600 rue University, Montreal, Quebec, H3A 2T8, Canada}

\author[0000-0001-8665-5523]{John J.~Ruan}
\affil{McGill Space Institute and Department of Physics, McGill University, 3600 rue University, Montreal, Quebec, H3A 2T8, Canada}

\author[0000-0001-6803-2138]{Daryl Haggard}
\affil{McGill Space Institute and Department of Physics, McGill University, 3600 rue University, Montreal, Quebec, H3A 2T8, Canada}
\affil{CIFAR Azrieli Global Scholar, Gravity \& the Extreme Universe Program, Canadian Institute for Advanced Research, 661 University Avenue,
Suite 505, Toronto, ON M5G 1M1, Canada}

\author[0000-0002-8465-3353]{Phil A. Evans}
\affil{Leicester Institute for Space and Earth Observation and Department of Physics \& Astronomy, University of Leicester, University Road, Leicester, LE1 7RH, UK}

\begin{abstract}
The multi-wavelength electromagnetic afterglow from the binary neutron star merger GW170817/GRB170817A has displayed long-term power-law brightening, and presented challenges to post-merger models of the non-thermal emission. The most recent radio observations up to 200 days post-merger suggest that the afterglow has finally peaked and may now be fading, but fading has not been confirmed in the X-rays. We present new, deep \Chandra\ observations of GW170817/GRB170817A at 260 days post-merger that reveal an X-ray flux of $F_\mathrm{0.3-8~keV} = 1.1\times10^{-14}$ erg s$^{-1}$ cm$^{-2}$, and confirm that the X-ray light curve is now also fading. Through rigorous comparisons to previous \Chandra\ observations of GW170817/GRB170817A, X-ray fading is detected between 160 and 260 days post-merger at a 4.4$\sigma$ significance, based on the X-ray data alone. We further constrain the X-ray photon index to steepen by $<$0.5 at 3.1$\sigma$ significance during this period, which disfavors the passing of the synchrotron cooling frequency through the X-ray band as the cause of the observed fading. These observations remain consistent with optically thin synchrotron afterglow emission. If this afterglow emission arises from a quasi-spherical mildly relativistic outflow, the X-ray fading suggests that the outflow is now decelerating. Alternatively, if this afterglow arises from a successful off-axis structured jet, the X-ray fading suggests that emission from the jet core has already entered the line of sight.

\end{abstract}

\section{Introduction}
\label{sec:intro}

The discovery of multi-wavelength electromagnetic emission from the binary neutron star (NS) merger GW170817 heralded the dawn of multi-messenger gravitational wave astronomy \citep[e.g.,][]{abbott17a, abbott17b, coulter17, evans17, goldstein17, hallinan17, soares17, troja17, valenti17}. The short Gamma-ray burst (sGRB) GRB170817A associated with this gravitational wave event confirmed that the progenitors of at least some sGRBs can be binary NS mergers \citep[BNS,][]{abbott17c, goldstein17, savchenko17}. Furthermore, the optical and infrared transient associated with GW170817 confirmed that the ejecta from binary neutron star mergers are the sites of \emph{r}-process nucleosythesis, in broad agreement with predictions from kilonova models \citep{arcavi17, cowperthwaite17, drout17, kasen17, kasliwal17, mccully17, pian17, shappee17, smartt17}. However, nearly nine months after its first detection, the post-merger evolution of this binary NS merger remains unclear, and the non-thermal electromagnetic emission is still rapidly evolving.

The early X-ray and radio light curves of GW170817/ GRB170817A are unlike any other sGRB previously observed. Early X-ray observations of the electromagnetic counterpart resulted in only upper limits on the X-ray flux \citep{evans17,margutti17}. An initial \Chandra\ X-ray detection was first made at $\sim$9 days post-merger \citep{troja17} and was confirmed in additional \Chandra\ observations at $\sim$15 days \citep{haggard17}. Similarly, early radio observations resulted in non-detections \citep{alexander17} until a first detection at $\sim$$16$ days post-merger \citep{hallinan17}. This delayed rise of the X-ray and radio emission is not observed in classical sGRB afterglows, which display monotonic fading over timescales of days \citep{fong17}. The X-ray and radio emission of GRB170817A was instead initially suggested to be consistent with models of a synchrotron afterglow from a simple top-hat sGRB jet observed off-axis, or a simple mildly relativistic cocoon blast-wave \citep[e.g.,][]{alexander17,haggard17,hallinan17,margutti17,troja17}.

Continued long-term X-ray and radio monitoring of GRB170817A has presented new challenges to post-merger models. The sky proximity of GRB170817A to the Sun prevented X-ray monitoring between $\sim$16 and $\sim$109 days post-merger. Meanwhile, continued radio monitoring revealed that GRB170817A continued to slowly brighten following a $t^{0.8}$ power-law over time up to $\sim$107 days post-merger \citep{mooley18a}. The slow long-term brightening of the afterglow emission now disfavors models of top-hat off-axis jets or simple cocoon blast-waves, which predict steeper power-law brightening than has been observed. This conclusion was strengthened by \Chandra\ X-ray observations at $\sim$109 days (immediately after Sun restrictions were lifted), which revealed that the X-ray emission brightened at a similar rate \citep{ruan18}. More recent X-ray observations at $\sim$160 days suggested that the afterglow light curve may be peaking \citep{davanzo18, margutti18}, and radio observations up to $\sim$200 days indicated that the afterglow may have begun fading \citep{dobie18}.

It is still unclear how the synchrotron afterglow emission arises during the post-merger evolution of GRB170817A. The consistency between the observed radio spectral index, X-ray spectral index, and broadband radio-to-X-ray spectral index (all $\alpha$ $\sim$ 0.6) conclusively shows that the spectral energy distribution of GRB170817A is a single $F_\nu \sim \nu^{-0.6}$ power-law that spans from X-ray through radio frequencies \citep[e.g.,][]{mooley18a, margutti18, ruan18}. This spectrum is consistent with optically-thin synchrotron emission from a trans-relativistic shock with a Lorentz factor of $\Gamma \approx 3 - 10$ \citep{lyman18,margutti18}. The constant slope of the synchrotron power-law spectrum between radio and X-ray frequencies up to 160 days implies that the synchrotron self-absorption frequency is below $\sim$1 GHz, while the synchrotron cooling frequency is above $\sim$1 keV.

Currently, the most promising models for the synchrotron afterglow invoke interactions between the relativistic jet and the merger ejecta, and can be crudely divided into (1) quasi-spherical, mildly relativistic outflows, and (2) off-axis angularly-structured jets. In outflow models, either dynamical ejecta or a cocoon shocked by a choked jet drives a mildly relativistic afterglow shock into the surrounding interstellar medium \citep[e.g.,][]{gottlieb17,gottlieb18, lazzati17b, mooley18a, nakar17}. This shock can be approximated as a quasi-spherical blast-wave, which will accelerate electrons that produce synchrotron emission in the shock-generated magnetic field. If the outflow is radially-stratified, such that the majority of the kinetic energy is in the lower-velocity material, the blast-wave experiences a continuous injection of energy in its coasting phase. The resulting afterglow emission will thus slowly brighten, as seen in the observations, particularly in comparison to simple non-stratified blast-waves \citep{mooley18a}. In contrast, for structured jet models the relativistic jet successfully breaks out of the ejecta, but is viewed off-axis \citep[e.g.][]{kath18, lamb17, lazzati17a,lyman18, margutti18}. The key feature of structured jet models is that the jet has angular structure, where the Lorentz factor $\Gamma$ of the jet decreases gradually as a function of angle from the jet axis, possibly due to interaction with kilonova ejecta \citep{lazzati17a,xie18}. When viewed off-axis, the observed synchrotron afterglow is dominated by jet material with increasingly larger initial $\Gamma$ over time, as the jet decelerates and material that was more initially relativistic enters the line of sight. This causes the structured jet afterglow to brighten slowly, similar to the observed X-ray and radio light curves, especially in comparison to jets with a top-hat distribution in $\Gamma$. Current multi-wavelength observations of the afterglow of GRB170817A during its brightening cannot distinguish between structured jets and outflows \citep[e.g.,][]{margutti18}, although some models of these two scenarios have predicted divergent afterglow light curve characteristics during the fading after its light curve peak \citep{lyman18, troja18, lamb18}. Furthermore, recent VLBI imaging suggest that the radio morphology of GRB170817A supports a successful off-axis angularly-structured jet \citep{mooley18b}.

Breaks in the afterglow light curve -- such as a light curve peak or a change in the light curve power-law slope -- place important constraints on models for the origin of its non-thermal emission. For an off-axis structured jet, a peak in the afterglow light curve will occur when emission from the jet core enters the line of sight \citep[e.g.,][]{vaneerten10, lyman18}. This geometric jet break is expected to be achromatic (occurring at all wavelengths simultaneously), and the timing of the break jointly constrains parameters such as the jet opening angle and the jet axis angle from the line of sight. For a quasi-spherical outflow with radially-stratified kinetic energy, a peak in the afterglow light curve will occur at the onset of deceleration of the slowest-moving material in the outflow shock \citep[e.g.,][]{nakar17}. This light curve break is also expected to be achromatic, and the timing of the break jointly constrains parameters such as the kinetic energy structure and Lorentz factor of the outflow. However, model interpretations and constraints based on an observed light curve break must first rule out other possible origins for the break.

The passing of the synchrotron cooling frequency $\nu_c$ through the X-ray band can also cause a beak in the X-ray light curve. For shock-accelerated relativistic electrons in the slow-cooling regime with a power-law distribution of energies, $\nu_c$ corresponds to the frequency above which electrons have now radiatively cooled. Since synchrotron radiative losses scale with particle energy, $\nu_c$ will decrease in frequency over time, crossing the X-ray band first before affecting the radio. The resulting characteristic steepening of the power-law spectrum across the cooling frequency of $\Delta$$\Gamma_\mathrm{X} = 0.5$ (where $\Gamma_\mathrm{X}$ is the X-ray photon index) causes a chromatic light curve break in which the X-ray light curve will fade before the radio. Observations up to 160 days post-merger show that the afterglow of GRB170817A continues to display a single power-law spectral energy distribution that spans from X-ray to radio frequencies \citep[e.g.,][]{ruan18, lyman18, margutti18}. This implies that $\nu_c$ was still above X-ray frequencies at that time, though it will eventually pass through the \Chandra\ band.

The exact timing of when $\nu_c$ is observed to cross the X-ray band is model dependent, and sensitive to model parameters. For example, in outflow models, $\nu_c(t)$ is most strongly dependent on the outflow velocity \citep{mooley18a}. The lower-velocity dynamical ejecta outflow model of \citet{hotokezaka18} predicts that the $\nu_c$ will cross the X-ray band on timescales of a few months to a year post-merger, while higher-velocity cocoon outflow models predict longer timescales of several years \citep{mooley18a}. Simulations of structured jets have predicted that $\nu_c$ will stay above X-ray frequencies for several years \citep{lazzati17a,margutti18}. Thus, a potential detection of a synchrotron cooling break through the X-rays in the near future would support the dynamical ejecta outflow model for the afterglow, while a non-detection would support either cocoon outflow models or structured jets. In any case, if a peak or break in the X-ray light curve is detected, interpretations for the origin of the break should first rule out synchrotron cooling as the cause of the break.

Although radio observations up to 200 days post-merger hinted that the afterglow of GRB170817A may have begun fading \citep{dobie18}, it has been unclear if this fading is also observed in X-rays. A recent \Chandra\ detection at 160 days suggest that the X-rays light curve is peaking \citep{margutti18}, but fading has yet to be confirmed. In this Letter, we present new, deep \Chandra\ X-ray observations of GRB170817A at 260 days post-merger, the first since the last \Chandra\ observation at 160 days. Our analysis of these new data and comparisons to previous \Chandra\ observations reveal that the X-ray emission is now also fading \citep[see also][]{alexander18}. Furthermore, we do not detect the characteristic steepening of the X-ray photon index expected from a synchrotron cooling break, thus disfavoring this possibility for the origin of the X-ray fading. In Section \ref{sec:observations}, we describe our new data and analysis procedure. In Section \ref{sec:discussion}, we compare our newest observations to previous \Chandra\ observations of GRB170817, to test for fading and changes in the X-ray photon index. We briefly conclude in Section \ref{sec:conclusion}.

\begin{figure} [t!]
\center{
\includegraphics[scale=0.4, angle=0.0]{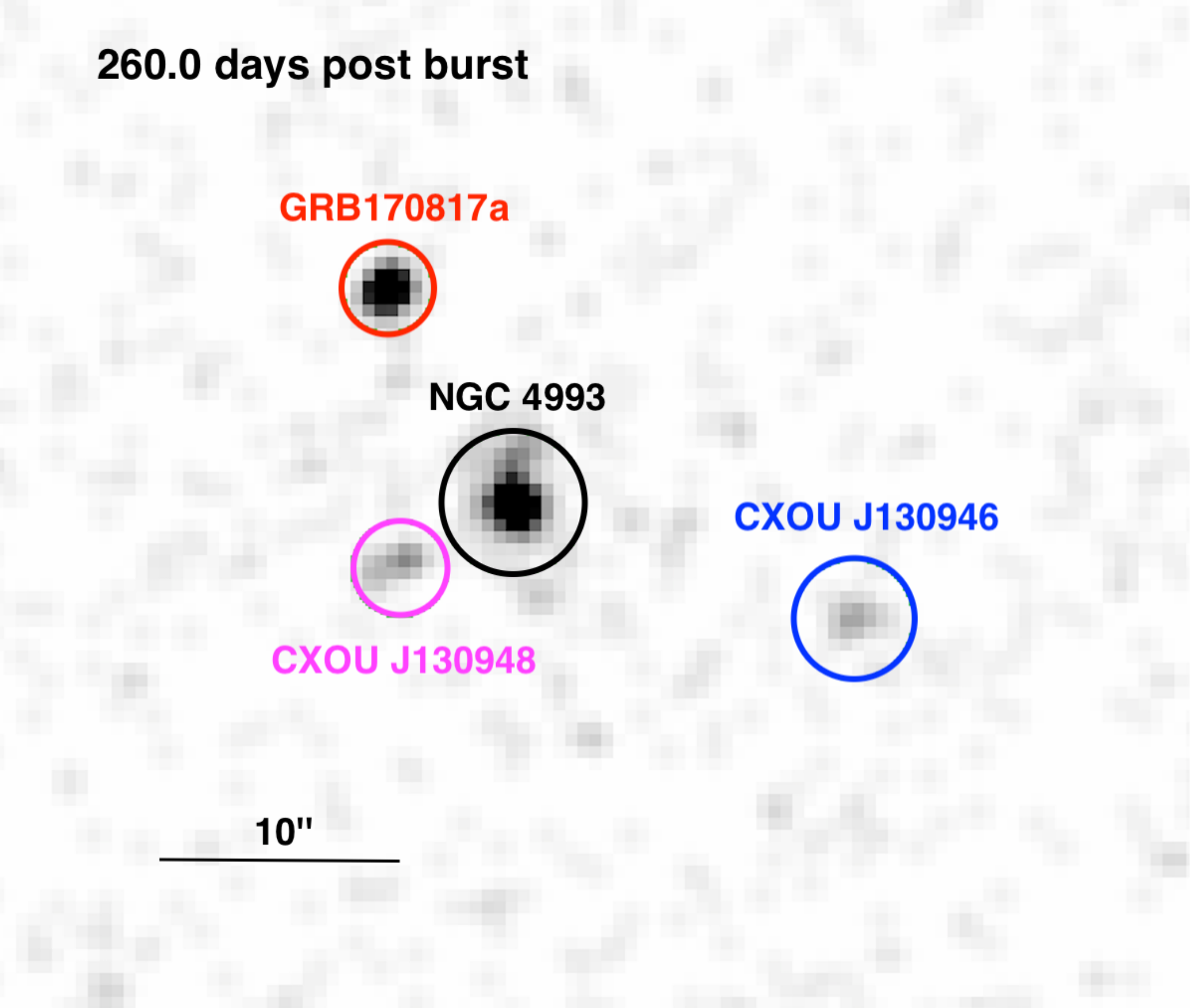}
}
\figcaption{The latest \Chandra\ $0.5-8.0$ keV X-ray image of GRB170817A at 260 days post-merger, in a 96.8~ks total observation. The X-ray afterglow of GW170817/GRB170817A is still clearly detected, along with X-ray emission from the host galaxy (NGC~4993) and two other previously-detected sources in the field. This image is shown on a linear scale, and has been smoothed with a 2-pixel Gaussian kernel.}
\label{fig:image}
\end{figure}

\section{X-ray Observations and Analysis}
\label{sec:observations}

\subsection{New \Chandra\ Observations \\ at 260 Days Post-Merger}
\label{ssc:newobs}
We have obtained new, deep \Chandra\ X-ray observations of GRB170817A via a Director's Discretionary Time allocation (PI: Wilkes, Program Number 19408644). Two exposures of GRB170817A were collected: (1) a 50.8 ks exposure (ObsID 21080) beginning at 2018 May 3.08 UT, approximately 259 days post-merger, and (2) a 46.0 ks exposure (ObsID 21090) beginning at 2018 May 4.08 UT, approximately 261 days post-merger. These two exposures where acquired using the ACIS-S3 chip in VFAINT mode. We use the CIAO v.4.10 software \citep[CALDB v4.7.8;][]{fruscione06} to reduce and analyze these \Chandra\ data. We first use CIAO's {\tt repro} script to reprocess all level 2 events files and apply the latest calibrations. To produce a deep X-ray image, we co-add the two exposures into a single 96.8 ks exposure at 260.0 days post-merger. The X-ray emission from GRB170817A is not expected to vary significantly over the 2-day timescales covered by these two exposures. The co-added 0.5--8 \Chandra\ keV image of GRB170817A at 260 days is shown in Figure \ref{fig:image}. X-ray emission is still clearly detected at the position of GRB170817A in these latest data, as well as at the positions of the three other nearby X-ray sources observed previously: CXOU~J130948, CXOU~130946, and the host galaxy NGC~4993.



We determine the centroid position of GRB170817A in each of the two individual $0.5-7$~keV images using the {\tt wavdetect} detection algorithm. We then extract X-ray spectra using regions with radii of 1\secspt97. This extraction radius corresponds to a $\sim$90\% encircled energy fraction near the \Chandra\ on-axis position. To determine the background, we use a large region from the same chip that does not overlap the extraction region of the detected sources.

We extract X-ray spectral and response files from the two individual observations of GRB170817A
using the {\tt specextract} tool, and co-add them into a single spectrum using 
{\tt combinespectra} to improve statistics. We use XSPEC v12.9.0 \citep{arnaud96} to fit the co-added spectra, with atomic cross sections from \citet{verner96} and abundances from \citet{wilms00}. For each source, we assume an absorbed power-law spectral model {\tt tbabs*powerlaw}  with fixed absorption $N_\mathrm{H} = 7.5\times10^{20}$ ~cm$^{-2}$. 
A distance of $42.5$~Mpc is adopted throughout, obtained from the host galaxy NGC~4993 \citep[z=0.0098 $D_L=42.5\pm0.3$][]{daCosta98}. 

\begin{figure*}[t!]
\center{
\includegraphics[scale=0.3,angle=0]{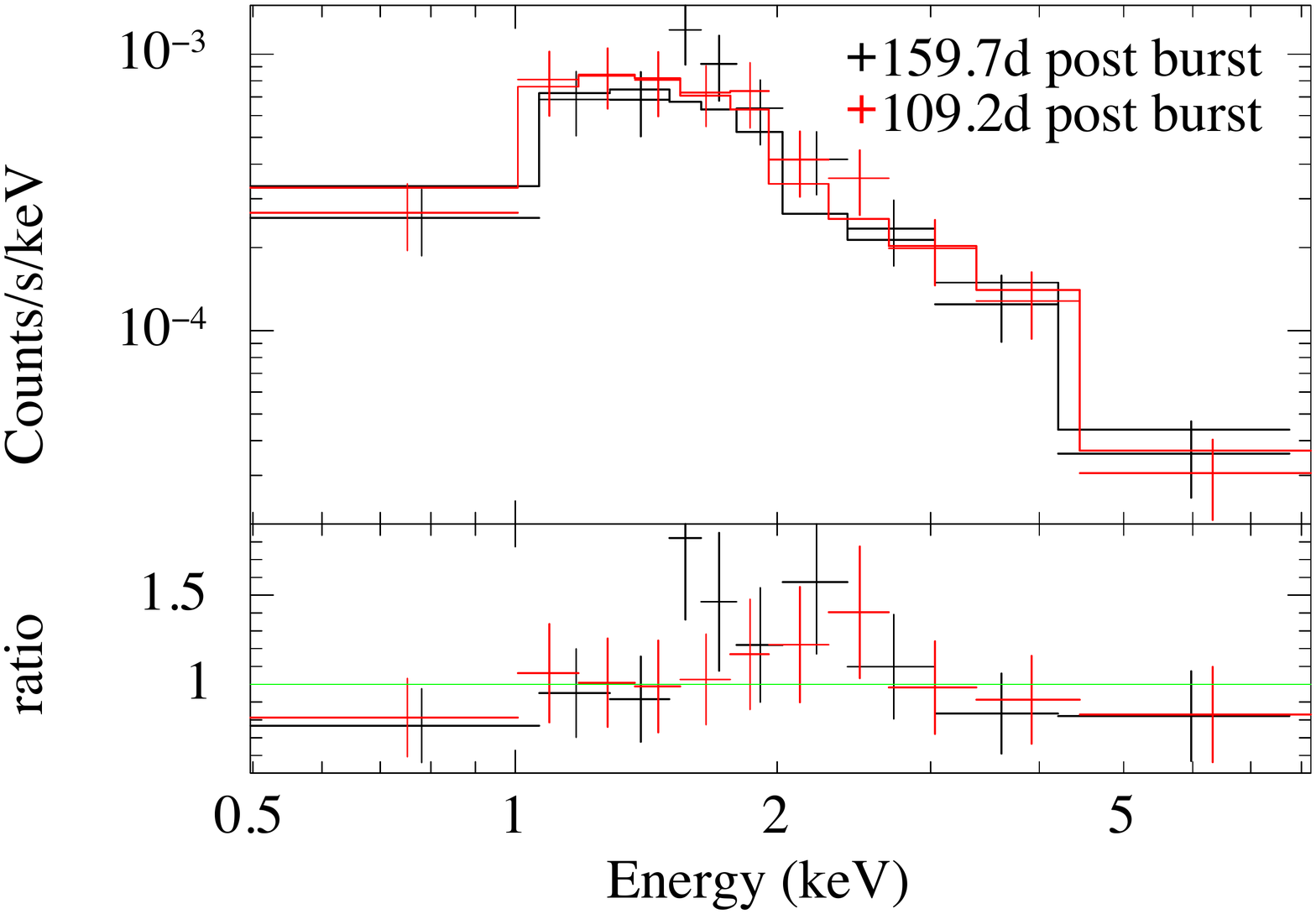}
\includegraphics[scale=0.3,angle=0]{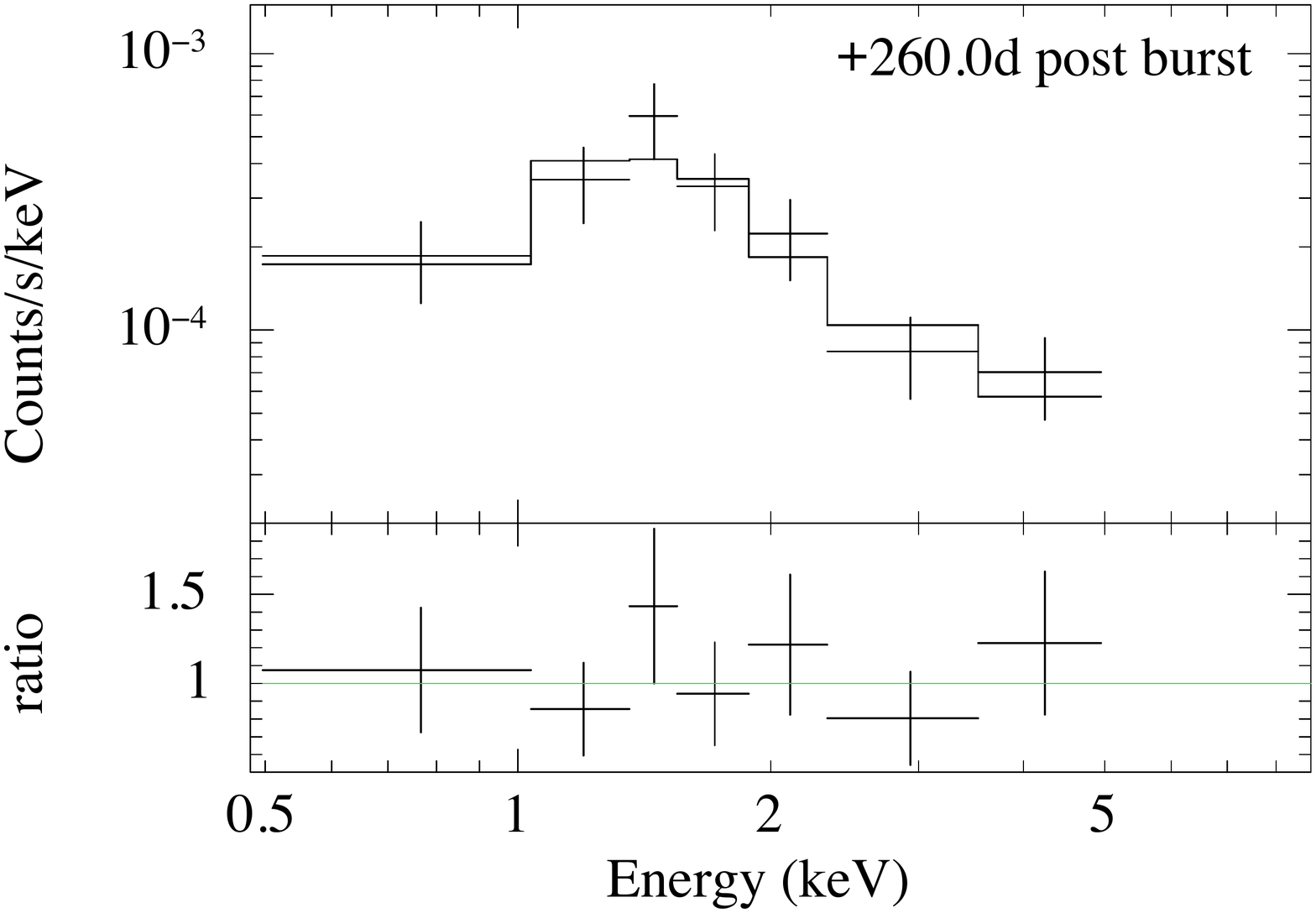}
}
\figcaption{ \Chandra\ X-ray spectra of GW170817/GRB170817A in the latest 96.8~ks observation at 260 days (right panel), compared to spectra from the previous $\sim$100~ks observations at 109 and 160 days (left panel). The best-fit absorbed power-law spectral models (solid lines) are jointly fit to their respective datasets: at 109 days $\Gamma_\mathrm{X} = 1.53^{+0.24}_{-0.23}$, at 160 days $\Gamma_\mathrm{X} = 1.58^{+0.23}_{-0.22}$, and at 260 days $\Gamma_\mathrm{X} = 1.57^{+0.38}_{-0.39}$.
The neutral hydrogen absorption column is fixed to $N$\textsubscript{H}$=7.5 \times10^{20}$~cm$^{-2}$ (see Table \ref{tab:fluxes} and Section \ref{sec:observations} for details). The lack of evolution in the spectral index disfavors the passing of the synchrotron cooling frequency through the X-ray band over the time interval probed to-date.
}
\label{fig:spec}
\vspace{0.2cm}
\end{figure*}

In this co-added 96.8 ks observation at 260 days, we measure GRB170817's source count rate to be $7.9\times10^{-4}$~counts~s$^{-1}$ ($0.5-8$~keV), a factor of $\sim$2 fainter than in the 104.9 ks \Chandra\ observation at 160 days. 
This count rate at 260 days corresponds to an absorbed flux of $F_\mathrm{0.3-8~keV} = 10.9 \times 10^{-15}$ erg s$^{-1}$ cm$^{-2}$, and an unabsorbed luminosity of $L_\mathrm{0.3-10~keV} = 2.96\times10^{39}$~erg~s$^{-1}$. The extracted X-ray spectrum of GRB170817A is shown in Figure \ref{fig:spec} (right panel) along with the best-fit spectral model. The spectrum is well-described by the assumed absorbed power-law model, with $\chi^2_{\nu}=0.89$. Table \ref{tab:fluxes} lists the best-fit power-law photon index and absorbed $0.3-8$~keV flux, in comparison to previous observations.

\begin{figure*}[t!]
\center{
\includegraphics[scale=0.7]{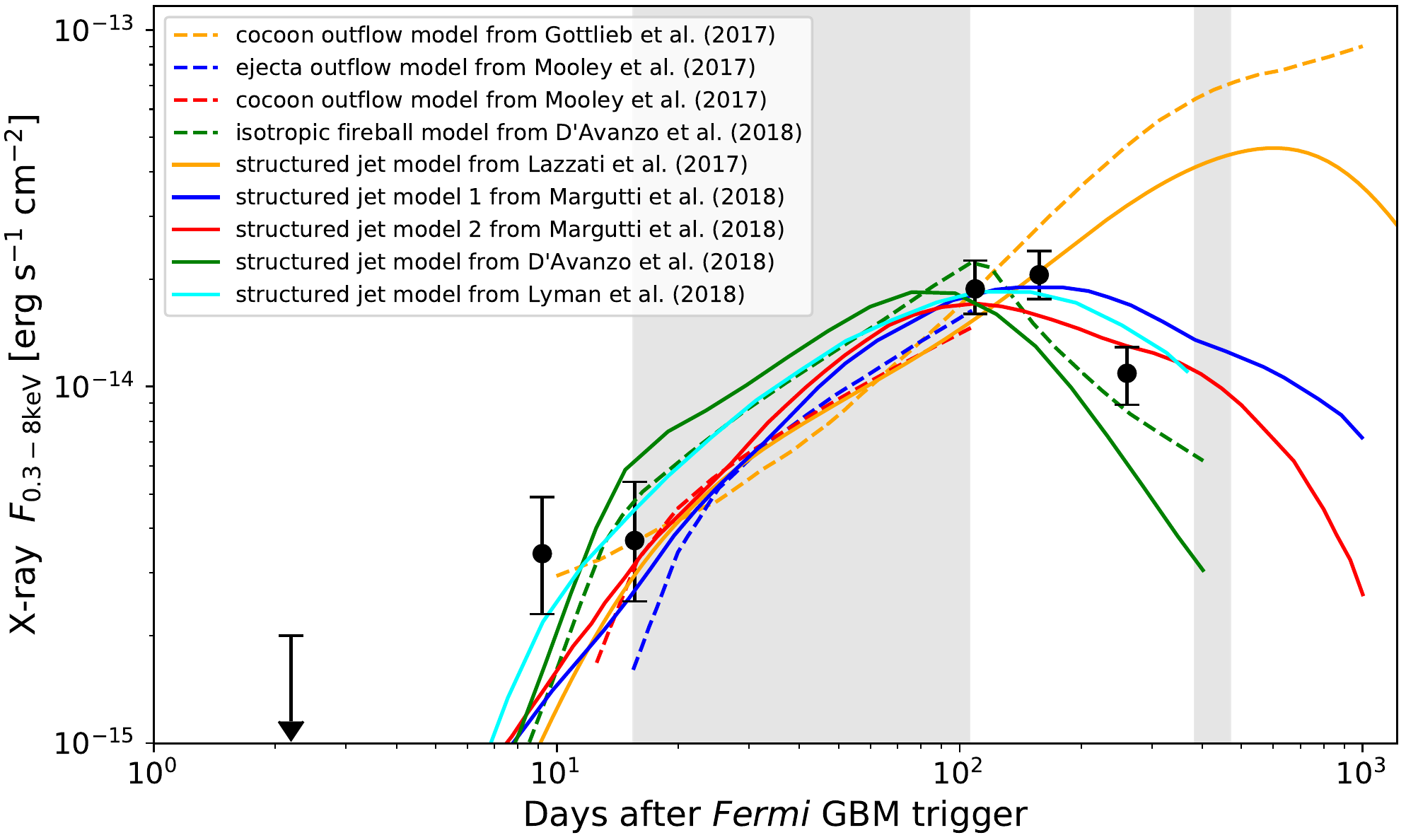}
}
\figcaption{\Chandra\ X-ray light curve of GW170817/GRB170817A (black points),
including the new observations at 260 days post-merger. The fluxes for all previous \Chandra\ observations are from a uniform re-reduction of the data (see Section \ref{ssc:prevobs}). X-ray light curve predictions from a selection of quasi-spherical outflow models (dashed lines) and structured jet models (solid lines) are also shown for comparison. We note that these models are fitted or matched to various combinations of previous X-ray/radio observations, and have flexible parameters that can produce a wide range of light curve peak times. Thus, the fading of the X-ray light curve revealed by our latest data at 260 days does not necessarily rule out any of the models shown. The gray shaded regions are time-spans over which \Chandra\ observations are not possible due to Sun constraints. All uncertainties shown are 90\% confidence level, and upper limits are 5$\sigma$ confidence.}
\label{fig:lightcurve}
\end{figure*}

\begin{deluxetable*}{lccccccccc}
\tablecaption{\Chandra\ X-ray Properties of GW170817/GRB170817A}
\tablehead{  
  \colhead{ObsID} & \colhead{PI} & \colhead{Start Date} & \colhead{Exposure} &  \colhead{Days Post-Merger} &  \colhead{Count Rate\tablenotemark{a}} & \colhead{Flux\tablenotemark{b}} & \colhead{Photon Index}& \colhead{Luminosity\tablenotemark{c}}\\
  \colhead{} & \colhead{} & \colhead{} & \colhead{[ks]} & \colhead{} & \colhead{[$10^{-4}$ cts s$^{-1}$]} & \colhead{[$10^{-14}~$erg~s$^{-1}$~cm$^{-2}$]} &  \colhead{$\Gamma_\mathrm{X}$} & \colhead{[$10^{38}$~erg~s$^{-1}$]} 
  }
\decimals
\startdata
 18955  & Fong  & 2017-08-19 & 24.64 & 2.3 & <1.2 & $<$0.13~\tablenotemark{d} & & <3.2 
\vspace{3pt}\\
\hline
19294  &  Troja & 2017-08-26 & 49.41 & 9.2 & $2.8\pm0.8$ & 0.34$^{+0.15}_{-0.11}$ & 1.6  & $9.2^{+4.6}_{-4.9}$ 
\vspace{3pt}\\
\hline
20728  &  Troja & 2017-09-01  & 46.69 & 15.4 \rdelim\}{2.2}{0pt}[15.6] & \multirow{2.2}{*}{$3.2\pm0.6$} &  \multirow{2.2}{*}{$0.36^{+0.17}_{-0.12}$} & \multirow{2.2}{*}{$2.42^{+0.95}_{-0.88}$} & \multirow{2.2}{*}{$10.8^{+5.2}_{-2.6}$}
\vspace{3pt}\\
18988  &  Haggard & 2017-09-02 & 46.69 & 15.9 &  &  
\vspace{3pt}\\
\hline
20860  &  Wilkes & 2017-12-03 & 74.09 & 108.0 \rdelim\}{2.2}{0pt}[109.2]  & \multirow{2.2}{*}{$14.8\pm1.3$}  &  \multirow{2.2}{*}{$1.88^{+0.38}_{-0.28}$} &  \multirow{2.2}{*}{$1.53^{+0.24}_{-0.23}$} & \multirow{2.2}{*}{$51.0^{+8.2}_{-9.3}$} 
\vspace{3pt}\\
20861  &  Wilkes & 2017-12-06   & 24.74 & 111.1 &   &  
\vspace{3pt}\\
\hline
20936  &  Wilkes & 2018-01-17 & 31.75 & 153.5 \rdelim\}{5.85}{0pt}[159.7] & \multirow{5.85}{*}{$15.3\pm1.2$}  &  \multirow{5.85}{*}{$2.06^{+0.34}_{-0.30}$} &  \multirow{5.85}{*}{$1.58^{+0.23}_{-0.22}$} & \multirow{5.85}{*}{$55.3^{+12.9}_{-8.9}$}
\vspace{3pt}\\
20938  &  Wilkes & 2018-01-21 & 15.86 & 157.1  &  &  & 
\vspace{3pt}\\
20937  &  Wilkes & 2018-01-23 & 20.77 & 158.9  & 
\vspace{3pt}\\
20939  &  Wilkes & 2018-01-24 & 22.25 & 159.9 &  &  
\vspace{3pt}\\
20945	  &  Wilkes & 2018-01-28 & 14.22 & 163.7 &  &  
\vspace{3pt}\\
\hline
21080	  &  Wilkes & 2018-05-03 & 50.78 & 259.2  \rdelim\}{2.2}{0pt}[260.0] &  \multirow{2.2}{*}{$7.9\pm0.9$} & \multirow{2.2}{*}{$1.09^{+0.24}_{-0.20}$} & \multirow{2.2}{*}{$1.57^{+0.38}_{-0.39}$} & \multirow{2.2}{*}{$29.6^{+7.1}_{-6.5}$}
\vspace{3pt}\\
21090	  &  Wilkes & 2018-05-05 & 46.00 & 260.8 &  &  
\vspace{3pt}\\
\enddata
\tablecomments{All reported uncertainties represent $90\%$ confidence intervals. The neutral hydrogen absorption was frozen to $N$\textsubscript{H}$=7.5 
\times10^{20}$~cm$^{-2}$ for all spectral fits, based on NGC 4993's A$_{\rm V}=0.338$ \citep{schlafly11}.}
\tablenotetext{a}{0.5-8 keV absorbed.} 
\tablenotetext{b}{0.3-8 keV absorbed.} 
\tablenotetext{c}{0.3-10 keV unabsorbed, assuming a luminosity distance of 42.5 Mpc.}
\tablenotetext{d}{Rescaled from \citet{margutti17}.}
\label{tab:fluxes}
\end{deluxetable*}


\subsection{Uniform Re-Reduction of \\ Previous \Chandra\ Observations}
\label{ssc:prevobs}

To enable a consistent and uniform comparison of the latest \Chandra\ observations of GRB170817A 
to previous observations, we systematically re-reduce and analyze all currently-available \Chandra\ data 
using the exact same procedure as in Section \ref{ssc:newobs}. For these 13 observations (listed in Table \ref{tab:fluxes}), we group observations that are close in time into co-added observations at 2.3, 9.2, 15.6, 109.2, and 159.7 days. The \Chandra\ data from the observation at 2.3 days are not currently publicly-available, so we rescale the 5$\sigma$ upper limit on the flux from \citet{margutti17} to an absorbed flux in the 0.3--8~keV band. 
An updated \Chandra\ light curve of GRB170817A using these fluxes is shown in Figure \ref{fig:lightcurve}, and the spectral fits to the 109 day and 160 day observations are shown in Figure \ref{fig:spec} (left panel). The resultant count rates, fluxes, and model parameters are consistent with previously-reported values.
In Section \ref{sec:discussion}, we use this uniformly-reduced dataset to statistically test whether the X-ray emission from GRB170817A has faded in the latest observations, and whether the X-ray photon index has steepened as expected for a synchrotron cooling break.

\begin{figure*}[t!]
\center{
\includegraphics[scale=0.6,angle=0]{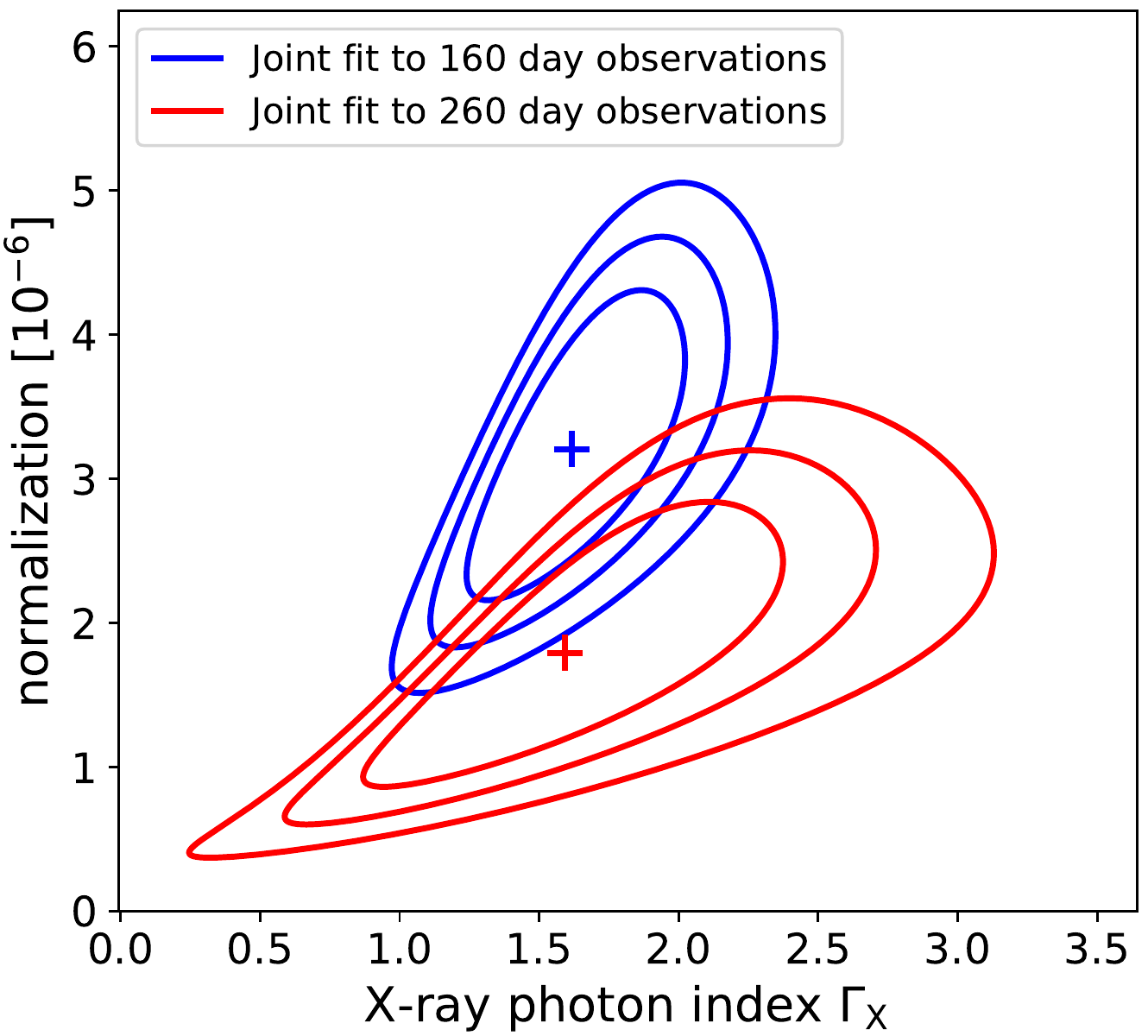}
\includegraphics[scale=0.6,angle=0]{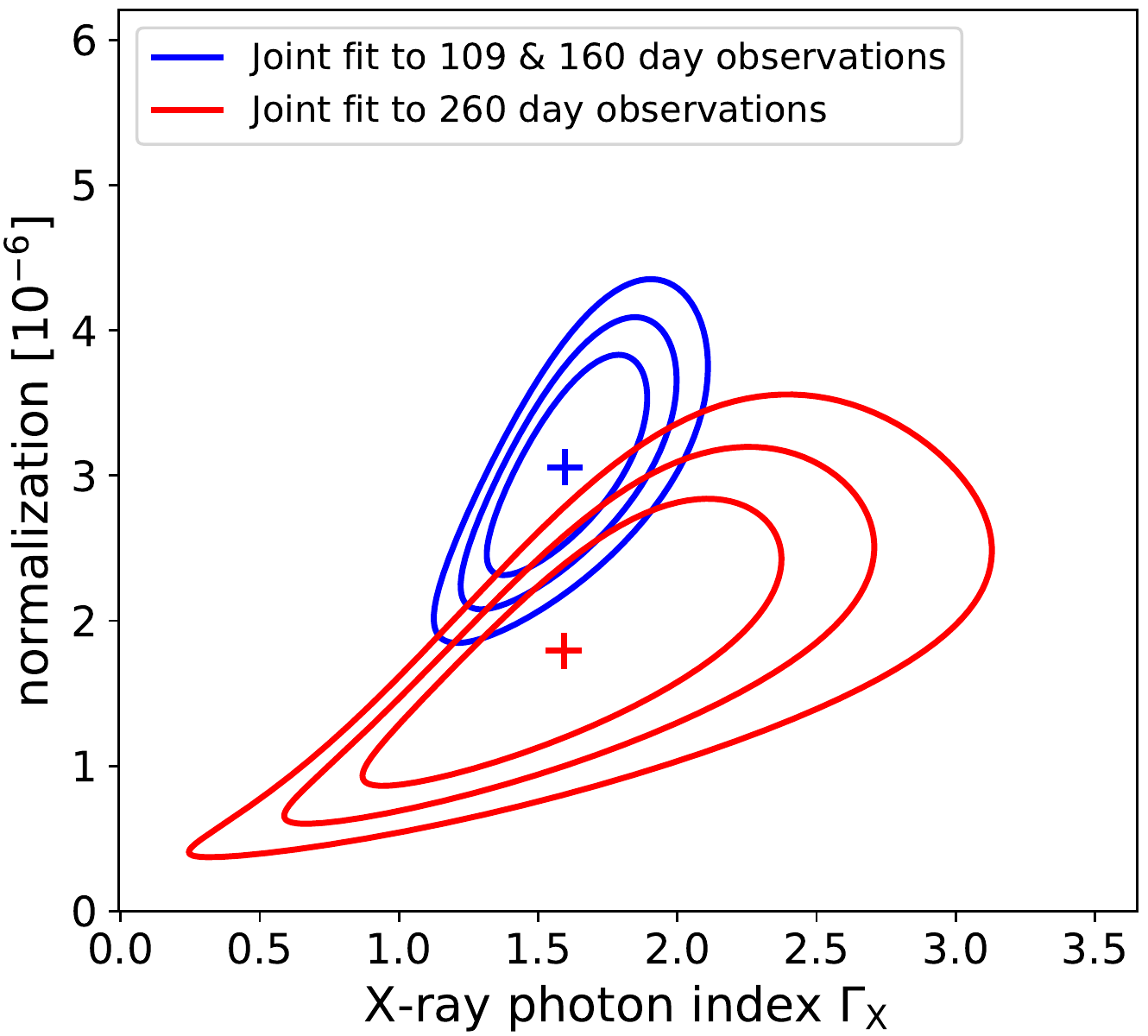}
}
\figcaption{Left: Comparison of the joint constraints on the X-ray photon index $\Gamma_\mathrm{X}$ and the flux normalization of the power-law X-ray spectrum of GRB170817A, between the 160 day \Chandra\ observations (blue contours) and the new 260 day observations (red contours). The contours show the 1$\sigma$, 90\%, and 2$\sigma$ confidence levels, and the best-fitting parameters are shown with a cross. We constrain the X-ray emission to be fading between these two epochs (i.e. with normalization decreasing) at a 4.4$\sigma$ significance, by marginalizing over $\Gamma_\mathrm{X}$ (see Section \ref{ssc:normchange}). Right panel: Similar joint constraints, comparing between the jointly-fitted 109 and 160 day data to the new 260 day observations. We constrain the X-ray photon index to steepen by $<$0.5 at a 3.1$\sigma$ significance by marginalizing over the normalization parameter, thus disfavoring a synchrotron cooling break as the cause of the fading (see Section \ref{ssc:gammachange}).}
\label{fig:contours}
\end{figure*}


\section{Comparisons to Previous Data}
\label{sec:discussion}

\subsection{Did the X-ray emission fade?}
\label{ssc:normchange}

We test whether the latest \Chandra\ observations at 260 days display statistically significant fading in comparison to the previous observation at 160 days.  Comparison of the observed $2-8$~keV count rates at 160 days and 260 days from Table 1 shows a decrease with 4.9$\sigma$ significance. However, this simplistic approach is useful only with the assumption that the spectral shape remains identical between the observations.  Similarly, the X-ray fluxes also cannot be directly compared to test for fading for two reasons.
First, there is a non-zero covariance between the power-law normalization $N$ and slope $\Gamma_\mathrm{X}$ parameters in the spectral power-law model $N\times E^{-\Gamma_\mathrm{X}}$ \citep{arnaud96}.  The normalization is defined as the intensity (ph~s$^{-1}$~cm$^{-2}$~keV$^{-1}$) of the source at 1~keV.  Integrated flux (with units of erg~s$^{-1}$~cm$^{-2}$ is dependent on both these parameters and 
thus, a change in the measured fluxes in Table 1 could be due to a change in $N$ and/or $\Gamma_\mathrm{X}$. Second, the uncertainties on both $N$ and $\Gamma_\mathrm{X}$ are non-Gaussian, and may have long tails that produce a large apparent change in the measured flux. Our approach is to marginalize the joint likelihood $\mathcal{L}$($N$, $\Gamma_\mathrm{X}$) at these two epochs over $\Gamma_\mathrm{X}$, and then compare the marginalized likelihood $\mathcal{L}$($N$) to test for a decrease in the power-law normalization between 160 and 260 days.

We first produce a joint likelihood $\mathcal{L}$($N$, $\Gamma_\mathrm{X}$) for both the 160 and 260 day observations. We use XSPEC to generate a 1000x1000 grid of $\chi^2$ as a function of both $\Gamma_\mathrm{X}$ and $N$, shown in Figure \ref{fig:contours} (left panel). We then use a exp($-\chi^2/2$) likelihood function to compute the joint likelihood $\mathcal{L}$($N$, $\Gamma_\mathrm{X}$). We marginalize over $\Gamma_\mathrm{X}$ to produce a marginalized $\mathcal{L}$($N$), for both the 160 and 260 day observations.

To test for a decrease in the power-law normalization $N$ (i.e., fading of the X-ray spectrum), we normalize the $\mathcal{L}$($N$) to produce a probability density function $p$($N$) for the 160 and 260 day observations. We then resample the two probability density functions to generate a distribution of $\Delta$$N$ between the two epochs. To perform the resampling of $p$($N$), we calculate the cumulative distribution function and use the inverse transform resampling method. We generate 10$^7$ resampled values for $N$ at both 160 and 260 days, and produce a distribution of $\Delta$$N$. Based on this $\Delta$$N$ distribution, we constrain the power-law spectrum normalization to be decreasing (i.e., $\Delta$$N < 0$) with 4.4$\sigma$ significance. Thus, our test shows that the X-ray emission from GRB170817A displays statistically significant fading between 160 and 260 days post-merger. Our conclusion based solely on X-ray data is consistent with alternative approaches to test for fading in the afterglow of GRB170817A based on joint fitting of X-ray, radio, and optical light curves \citep{alexander18}.


\subsection{Did the X-ray photon index steepen?}
\label{ssc:gammachange}
The statistically-significant fading of the X-ray emission from GRB170817A revealed in the latest \Chandra\ observations at 260 days can occur if the synchrotron cooling frequency $\nu_c$ has recently passed through the \Chandra\ band. In this scenario, the X-ray photon index $\Gamma_\mathrm{X}$ would be observed to steepen by a characteristic $\Delta\Gamma_\mathrm{X} = 0.5$, and the X-ray light curve would fade faster than the radio until $\nu_c$ later passes through radio frequencies. Thus, we test for synchrotron cooling as the origin of the X-ray light curve peak, by comparing the observed change in $\Gamma_\mathrm{X}$ to the characteristic steepening of $\Delta\Gamma_\mathrm{X} = 0.5$ expected from synchrotron cooling.

Similar to our test for X-ray fading in Section \ref{ssc:normchange}, a test for changes in $\Gamma_\mathrm{X}$ also should not be based on a direct comparison of the fitted $\Gamma_\mathrm{X}$ between two observations in Table 1. We thus follow a similar procedure as in Section \ref{ssc:normchange} to test for changes in $\Gamma_\mathrm{X}$, but now marginalizing $\mathcal{L}$($N$, $\Gamma_\mathrm{X}$) over the normalization $N$. Since X-ray fading was not observed between 109 and 160 days, we jointly-fit the observations between these two epochs (constraints shown in right panel of Figure \ref{fig:contours}, and best-fit spectral model shown in left panel of Figure \ref{fig:spec}), and compare this joint fit to the newest observations at 260 days. Based on the distribution in $\Delta\Gamma_\mathrm{X}$ between these two datasets, we constrain $\Gamma_\mathrm{X}$ to steepen by $<$0.5 at a 3.1$\sigma$ significance. Thus, our test disfavors the characteristic steepening of the power-law X-ray spectrum from a synchrotron cooling break as the cause of the X-ray fading between 160 and 260 days post-merger.


\subsection{A Best-Available Constraint \\ on the X-ray Photon Index}

We use our entire uniformly-reduced \Chandra\ dataset from Section \ref{ssc:prevobs} 
to derive the best constraints to-date for $\Gamma_\mathrm{X}$. Since the X-ray flux of GRB170817A changes over time, we cannot directly co-add all the observations, and instead use the co-added data from the groups of observations at 9.2, 15.6, 109.2, 159.7, and 260.0 days to jointly fit an absorbed power-law model. The power-law photon index $\Gamma_\mathrm{X}$ is tied between observations and the normalization left free. In our tests, the observations at 9.2 and 15.6 days do not improve the spectral fit due to low count rates.
We thus only use the observations at 109.2, 159.7, and 260.0 days, and derive a jointly-fitted $\Gamma_\mathrm{X} = 1.56^{+0.14}_{-0.15}$. This is the best constraint on $\Gamma_\mathrm{X}$ to-date and is highly consistent with the current best constraints on both the radio spectral index of $\alpha_\mathrm{R} = 0.61\pm0.05$ (where $\alpha = \Gamma - 1$) \citep{mooley18a}, and the broadband X-ray-to-radio spectral index of $\alpha_\mathrm{XR} = 0.585\pm0.005$ \citep{margutti18}. The consistency of the X-ray photon index with the radio spectral index further supports a synchrotron power-law spectrum that spans from X-ray to radio frequencies.


\section{Conclusion}
\label{sec:conclusion}

We present new, deep \Chandra\ observations of GW170817/GRB170817A at 260 days post-merger. These are the first X-ray observations since 160 days, which had suggested that the brightening afterglow light curve may be reaching a peak. Our analysis of the new observations reveals that GRB170817A is indeed now fading in X-rays. We compare the X-ray properties of GRB170817A to previous \Chandra\ observations using a uniform re-reduction of all available data. We show that the fading of the X-ray power-law spectrum is detected at a 4.4$\sigma$ significance, based on the X-ray data alone. Furthermore, we detect no change in the X-ray photon index $\Gamma_\mathrm{X}$, and constrain $\Gamma_\mathrm{X}$ to steepen by <0.5 at 3.1$\sigma$ significance. This disfavors a synchrotron cooling break as the cause of the X-ray fading, which would instead result in a characteristic steeping in $\Gamma_\mathrm{X}$ of 0.5. 

The X-ray fading of GRB170817A remains consistent with current post-merger models for the synchrotron afterglow. By jointly-fitting all available \Chandra\ data to date, we derive a best-available measurement of the X-ray photon index of $\Gamma_\mathrm{X} = 1.56^{+0.14}_{-0.15}$. The consistency of $\Gamma_\mathrm{X}$ with the radio spectral index implies that the afterglow spectrum remains a single power-law spanning from radio to X-ray frequencies. For quasi-spherical mildly relativistic outflow models of the afterglow emission, the fading at radio through X-ray frequencies implies that the outflow is now in a decelerating phase. For angularly-structured off-axis jets, the fading implies that emission from core of the jet has already entered the line of sight. Both \citet{lyman18} and \citet{troja18} suggest that the post-break light curve characteristics of GRB170817A can discriminate between the structured jet and quasi-spherical outflow models. To this end, we urge that continued X-ray monitoring of GRB170817A be avidly pursued.

Looking forward, future detections of electromagnetic counterparts to LIGO-Virgo gravitational wave sources will ideally have long, multi-wavelength light curves. Absent this ideal, our analysis indicates that deep X-ray observations alone are sufficient to track the post-merger evolution of this binary NS merger, and monitor the progression of both light curve breaks and potential spectral breaks as the outflow/jet evolves. If this finding is supported by future detections, it could provide an important constraint on X-ray mission design and electromagnetic follow-up strategies.


\acknowledgments
The authors thank Belinda Wilkes and the \Chandra\ scheduling, data processing, and archive teams for making these observations possible. This work was supported by \Chandra\ Award Number GO7-18033X, issued by the {\it Chandra X-ray Observatory Center}, which is operated by the Smithsonian Astrophysical Observatory for and on behalf of the National Aeronautics Space Administration (NASA) under contract NAS8-03060. J.J.R., M.N., and D.H. acknowledge support from a Natural Sciences and Engineering Research Council of Canada (NSERC) Discovery Grant and a Fonds de recherche du Qu\'{e}bec--Nature et Technologies (FRQNT) Nouveaux Chercheurs Grant. J.J.R. and M.N. acknowledge funding from the McGill Trottier Chair in Astrophysics and Cosmology. D.H. acknowledges support from the Canadian Institute for Advanced Research (CIFAR). PAE acknowledges UKSA support. 

\facility{CXO}
\software{Matplotlib \citep{hunter07}}
          

\bibliographystyle{apj}

\begin{thebibliography}{}

\bibitem[Abbott et al.(2017a)]{abbott17a} Abbott, B.~P., Abbott, R., Abbott, T.~D., et al.\ 2017a, Physical Review Letters, 119, 161101 

\bibitem[Abbott et al.(2017b)]{abbott17b} Abbott, B.~P., Abbott, R., Abbott, T.~D., et al.\ 2017b, \apjl, 848, L12 

\bibitem[Abbott et al.(2017c)]{abbott17c} Abbott, B.~P., Abbott, R., Abbott, T.~D., et al.\ 2017c, \apjl, 848, L13 

\bibitem[Alexander et al.(2017)]{alexander17} Alexander, K.~D., Berger, E., Fong, W., et al.\ 2017, \apjl, 848, L21 

\bibitem[Alexander et al.(2018)]{alexander18} Alexander, K.~D., Margutti, R., Blanchard, P.~K., et al.\ 2018, arXiv:1805.02870 

\bibitem[Arcavi et al.(2017)]{arcavi17} Arcavi, I., Hosseinzadeh, G., Howell, D.~A., et al.\ 2017, \nat, 551, 64 

\bibitem[Arnaud(1996)]{arnaud96} Arnaud, K.~A.\ 1996, Astronomical Data Analysis Software and Systems V, 101, 17 

\bibitem[Coulter et al.(2017)]{coulter17} Coulter, D.~A., Foley, R.~J., Kilpatrick, C.~D., et al.\ 2017, arXiv:1710.05452

\bibitem[Cowperthwaite et al.(2017)]{cowperthwaite17} Cowperthwaite, P.~S., Berger, E., Villar, V.~A., et al.\ 2017, \apjl, 848, L17 

\bibitem[da Costa et al.(1988)]{daCosta98} da Costa, L.~N., Pellegrini, P.~S., Sargent, W.~L.~W., et al.\ 1988, \apj, 327, 544 

\bibitem[D'Avanzo et al.(2018)]{davanzo18} D'Avanzo, P., Campana, S., Ghisellini, G., et al.\ 2018, arXiv:1801.06164 

\bibitem[Dobie et al.(2018)]{dobie18} Dobie, D., Kaplan, D.~L., Murphy, T., et al.\ 2018, arXiv:1803.06853 

\bibitem[Drout et al.(2017)]{drout17} Drout, M.~R., Piro, A.~L., Shappee, B.~J., et al.\ 2017, arXiv:1710.05443 

\bibitem[Evans et al.(2017)]{evans17} Evans, P.~A., Cenko, S.~B., Kennea, J.~A., et al.\ 2017, arXiv:1710.05437 

\bibitem[Fong et al.(2017)]{fong17} Fong, W., Berger, E., Blanchard, P.~K., et al.\ 2017, \apjl, 848, L23 

\bibitem[Fruscione et al.(2006)]{fruscione06} Fruscione, A., McDowell, J.~C., Allen, G.~E., et al.\ 2006, \procspie, 6270, 62701V 

\bibitem[Goldstein et al.(2017)]{goldstein17} Goldstein, A., Veres, P., Burns, E., et al.\ 2017, \apjl, 848, L14 

\bibitem[Gottlieb et al.(2017)]{gottlieb17} Gottlieb, O., Nakar, E., Piran, T., \& Hotokezaka, K.\ 2017, arXiv:1710.05896 

\bibitem[Gottlieb et al.(2018)]{gottlieb18} Gottlieb, O., Nakar, E., \& Piran, T.\ 2018, \mnras, 473, 576 

\bibitem[Haggard et al.(2017)]{haggard17} Haggard, D., Nynka, M., Ruan, J.~J., et al.\ 2017, \apjl, 848, L25 

\bibitem[Hallinan et al.(2017)]{hallinan17} Hallinan, G., Corsi, A., Mooley, K.~P., et al.\ 2017, arXiv:1710.05435 

\bibitem[Hotokezaka et al.(2018)]{hotokezaka18} Hotokezaka, K., Kiuchi, K., Shibata, M., Nakar, E., \& Piran, T.\ 2018, arXiv:1803.00599 

\bibitem[Hunter(2007)]{hunter07} Hunter, J.~D.\ 2007, Computing in Science and Engineering, 9, 90 

\bibitem[Kasen et al.(2017)]{kasen17} Kasen, D., Metzger, B., Barnes, J., Quataert, E., \& Ramirez-Ruiz, E.\ 2017, \nat, 551, 80 

\bibitem[Kasliwal et al.(2017)]{kasliwal17} Kasliwal, M.~M., Nakar, E., Singer, L.~P., et al.\ 2017, arXiv:1710.05436 

\bibitem[Kathirgamaraju et al.(2018)]{kath18} Kathirgamaraju, A., Barniol Duran, R., \& Giannios, D.\ 2018, \mnras, 473, L121 

\bibitem[Lamb \& Kobayashi(2017)]{lamb17} Lamb, G.~P., \& Kobayashi, S.\ 2017, \mnras, 472, 4953 

\bibitem[Lamb et al.(2018)]{lamb18} Lamb, G.~P, Mandel, I., \& Resmi, L.\ 2018, arXiv:1806.03843 

\bibitem[Lazzati et al.(2017a)]{lazzati17a} Lazzati, D., Perna, R., Morsony, B.~J., et al.\ 2017a, arXiv:1712.03237 

\bibitem[Lazzati et al.(2017b)]{lazzati17b} Lazzati, D., Deich, A., Morsony, B.~J., \& Workman, J.~C.\ 2017b, \mnras, 471, 1652 

\bibitem[Lyman et al.(2018)]{lyman18} Lyman, J.~D., Lamb, G.~P., Levan, A.~J., et al.\ 2018, arXiv:1801.02669 

\bibitem[Margutti et al.(2017)]{margutti17} Margutti, R., Berger, E., Fong, W., et al.\ 2017, \apjl, 848, L20 

\bibitem[Margutti et al.(2018)]{margutti18} Margutti, R., Alexander, K.~D., Xie, X., et al.\ 2018, \apjl, 856, L18 

\bibitem[McCully et al.(2017)]{mccully17} McCully, C., Hiramatsu, D., Howell, D.~A., et al.\ 2017, \apjl, 848, L32 

\bibitem[Mooley et al.(2018a)]{mooley18a} Mooley, K.~P., Nakar, E., Hotokezaka, K., et al.\ 2018a, \nat, 554, 207

\bibitem[Mooley et al.(2018b)]{mooley18b} Mooley, K.~P., Deller, A.~T., Gottlieb, O., et al.\ 2018b, arXiv:1806.09693 

\bibitem[Nakar \& Piran(2017)]{nakar17} Nakar, E., \& Piran, T.\ 2017, \apj, 834, 28 

\bibitem[Nakar \& Piran(2018)]{nakar18a} Nakar, E., \& Piran, T.\ 2018, arXiv:1801.09712 

\bibitem[Nakar et al.(2018)]{nakar18b} Nakar, E., Gottlieb, O., Piran, T., Kasliwal, M.~M., \& Hallinan, G.\ 2018, arXiv:1803.07595 

\bibitem[Pian et al.(2017)]{pian17} Pian, E., DAvanzo, P., Benetti, S., et al.\ 2017, \nat, 551, 67 

\bibitem[Ruan et al.(2018)]{ruan18} Ruan, J.~J., Nynka, M., Haggard, D., Kalogera, V., \& Evans, P.\ 2018, \apjl, 853, L4 

\bibitem[Savchenko et al.(2017)]{savchenko17} Savchenko, V., Ferrigno, C., Kuulkers, E., et al.\ 2017, \apjl, 848, L15 

\bibitem[Schlafly \& Finkbeiner(2011)]{schlafly11} Schlafly, E.~F., \& Finkbeiner, D.~P.\ 2011, \apj, 737, 103 

\bibitem[Shappee et al.(2017)]{shappee17} Shappee, B.~J., Simon, J.~D., Drout, M.~R., et al.\ 2017, arXiv:1710.05432 

\bibitem[Smartt et al.(2017)]{smartt17} Smartt, S.~J., Chen, T.-W., Jerkstrand, A., et al.\ 2017, \nat, 551, 75 

\bibitem[Soares-Santos et al.(2017)]{soares17} Soares-Santos, M., Holz, D.~E., Annis, J., et al.\ 2017, \apjl, 848, L16 

\bibitem[Troja et al.(2017)]{troja17} Troja, E., Piro, L., van Eerten, H., et al.\ 2017, \nat, 551, 71 

\bibitem[Troja et al.(2018)]{troja18} Troja, E., Piro, L., Ryan, G., et al.\ 2018, \mnras

\bibitem[Valenti et al.(2017)]{valenti17} Valenti, S., David, Sand, J., et al.\ 2017, \apjl, 848, L24 

\bibitem[van Eerten et al.(2010)]{vaneerten10} van Eerten, H., Zhang, W., \& MacFadyen, A.\ 2010, \apj, 722, 235 

\bibitem[Verner et al.(1996)]{verner96} Verner, D.~A., Ferland, G.~J., Korista, K.~T., \& Yakovlev, D.~G.\ 1996, \apj, 465, 487 

\bibitem[Wilms et al.(2000)]{wilms00} Wilms, J., Allen, A., \& McCray, R.\ 2000, \apj, 542, 914 

\bibitem[Xie et al.(2018)]{xie18} Xie, X., Zrake, J., \& MacFadyen, A.\ 2018, arXiv:1804.09345 

\end{thebibliography}

\end{document}